# Using Linked Micromaps to Explore Complex Structures in Official Statistics

**Randall Powers, Darcy Steeg Morris, John L. Eltinge and Wendy L. Martinez[1]**

## Abstract

Over the past decade, researchers have focused increasing levels of attention on the use of survey and non-survey data to inform decision-making by multiple stakeholders. Work with such data generally requires extensive exploration before a statistics practitioner focuses on specific steps in model building and inference. For many of the resulting initial exploratory analyses, crucial issues center on the extent to which empirical results may vary over geography and subpopulations. Such information is usually presented in tabular form, which can be difficult for stakeholders and decision makers to understand and to utilize.

To address these issues, this paper uses data from the U.S. Bureau of Labor Statistics to illustrate a suite of tools known as *linked micromaps*. These applications show how linked micromaps can help stakeholders better understand and view descriptive statistics for populations and subpopulations, explore multivariate relationships and ordinal structure, and discover patterns of heterogeneity across time and space. In addition, this paper comments briefly on the prospective use of linked micromaps in model-building and analysis of multiple components of uncertainty.

The R package called `micromapST` (available at CRAN https://cran.r-project.org/) was used to create the linked micromaps in this article. The R scripts and public data used to create the graphics are available at https://github.com/wendylmartinez/Linked-Micromaps-JSTP26/ for reproducibility and to learn more about this graphical approach.

[1]Randall Powers, Office of Survey Research Methods Research, U.S. Bureau of Labor Statistics; Darcy Steeg Morris, Center for Statistical Research and Methodology, U.S. Census Bureau; John L. Eltinge, Research and Methodology Directorate, U.S. Census Bureau (Retired); Wendy L. Martinez, Research and Methodology Directorate, U.S. Census Bureau (corresponding author: wendy.l.martinez@census.gov).

## 1. Introduction

### 1.1 Exploratory Analysis of Information Produced by National Statistical Offices

National Statistical Offices (NSOs), as well as academic and private-sector organizations, often produce large amounts of statistical information in the form of complex tables that are disseminated in reports, tabbed spreadsheet files, or data streams through application programming interfaces (APIs). Analyses of such information may involve overall measures of centrality (e.g., means and medians), but also focus on patterns of variability related to:

- Differences across groups defined by one or more levels of nested and cross-sectional structure based on characteristics such as temporal, geographic, demographic, industrial and occupational classification factors.
- Proximity and ordinal structure related to these groups.
- Variability across different measured outcome variables (e.g., employment, income, expenditures or health conditions).
- Multivariate relationships, as reflected in parametric or nonparametric regression models.
- Sampling variability and other sources of error.

Examination of such tables can be useful, but it can also be burdensome. Consequently, data owners often use visualization methods to help users see overall patterns more readily than they could by examining the large sets of corresponding tables. These methods can be especially useful when analysts seek to explore those patterns across multiple geographical areas or other subpopulations and to determine whether those patterns are roughly comparable across those areas. Examples of common visualization methods for spatially indexed data include those described in Section 1.3 below. Each of those methods has prospective strengths and limitations. This paper will focus on one class of visualization methods known as *linked micromaps* that can be especially useful in exploring the questions of variability described above. Some notable features of linked micromaps are illustrated in this paper using two large-scale publicly available data sets.

The remainder of Section 1 contains information on the data sets used in the article, as well as an introduction to linked micromaps. Section 2 explains how linked micromaps can be used to explore subpopulations, such as industries and occupations. Section 3 demonstrates the use of linked micromaps to understand multivariate relationships and to provide side-by-side comparisons of related statistics. Section 4 notes several prospective extensions of the ideas considered in this paper. This article focuses on data and examples for the 50 U.S. states and the District of Columbia. However, these ideas can be applied to other countries and geographies (see Figure 2.3).

## 1.2 Data Sets

The applications described in this paper use data from two publicly available tabular sources described below. For these sources, the tables include many variables based on geography, industry, occupations, and more. These rich and complex data structures can provide useful detailed information. However, the tables may be difficult to navigate and challenging for the reader to see overall patterns and relationships in the tables. These challenges may be especially acute for tables that include important information on variability across and within geographical areas and subpopulations. Subsequent sections will show how linked micromaps can address those concerns.

Our first data set is from the Quarterly Census of Employment and Wages (QCEW, https://www.bls.gov/cew) program, which is managed by the U.S. Bureau of Labor Statistics (BLS). The QCEW publishes information on employment and wages on a quarterly basis. It is available at the national, state, county, Metropolitan Statistical Area (MSA), and Balance of State (BOS) levels. QCEW produces data aggregated to many different industry levels. The QCEW has some motivating features for graphical exploration. The geographical and industrial structure of the QCEW is both hierarchical and complex. Also, the QCEW is a nominal census based on administrative data, and thus in principle is not subject to sampling error. An easy way to download data for specific industries and geographies is to use the BLS State and County Map tool found here https://data.bls.gov/maps/cew/us. Users can choose the time point (year/quarter), industry, and statistic, which are then viewed in a choropleth map and downloaded.

The second illustrative data set explored in this paper is the Occupational Employment and Wage Statistics (OEWS) program of the BLS that produces annual estimates of employment and wages for approximately 830 occupations (https://www.bls.gov/soc). Estimates are published for individual states and for metropolitan and non-metropolitan areas, as well as the national level. The OEWS is an establishment survey with both geographical structure and multivariate relationships that are of interest to analysts. The QCEW focuses on establishments, while the OEWS has data concentrating on occupations. An interactive tool to explore OEWS data is available here: https://data.bls.gov/oesmap. This tool produces choropleth maps for selected statistics or measures (location quotient, mean wages, employment, etc.) and occupations (major or detailed groups). Data tables for national, state, and sub-geographies are available for download at this site https://www.bls.gov/oes/tables.htm.

## 1.3 Common Ways to Visualize Spatial Data

Before exploring the features and benefits of linked micromaps, it is helpful to provide some context by reviewing other methods of displaying geographic data and describing their limitations. See Powers, et al. (2024) for details and examples.

Bar charts are often used to display state-level statistics, in part because they are easily created in Microsoft Excel and other statistical programs. Horizontal bars are typically used for US state-level data where the length of each bar represents the statistic of interest (e.g., total number of employees in a state, etc.). The bars are usually ordered alphabetically by state, so stakeholders can readily find the sub-region, but the distribution of the statistic is difficult, if not impossible, to understand. (For some related comments, see, e.g., Wainer, 1984, Rule 9.) One could also order the bars by the statistic, providing information on its distribution. In both cases, it is difficult or impossible to discern spatial dependencies and regional effects, the understanding of which might inform model building and inference.

Spatial data can be presented in a table and again ordered alphabetically by state or region. As with bar charts, this method of presentation makes it difficult to extract information, relationships, and insights regarding the connection between statistics and geography. It is up to the viewer to make a mental comparison of data throughout the table.

Choropleth maps are often used to display geographic data connected to a spatial unit, such as a state or county. In choropleth maps, the color indicates some variable of interest. Choropleth maps are very popular for good reasons. They are intuitive for most people to understand, and they show spatial distributions of one characteristic of data quite well. However, there are some limitations (Tukey, 1979; Carr and Pierson, 1996). It is sometimes difficult to distinguish values using color, especially when color gradients are used. There is no ordering of the statistics (and hence the geography), except through the legend. Choropleth maps give the less populated but large land mass states more prominence, making less populous states like Wyoming visually more distinct or important and tiny states like Rhode Island easy to miss.

An enhancement to choropleth maps is the bubble or proportional symbol map. This approach adds a bubble or circle over each region that represents one or two additional values using the size of the bubble and its color. These also have limitations (Robbins, 2013, Chapter 3). It can be hard to ascertain exact values using circle sizes and color scales, and the symbols can overlap, making it difficult to see patterns.

### 1.4 Linked Micromaps – A Brief Introduction

*Conceptual Basis.* Linked micromaps were originally developed by Carr and Pierson (1996) in response to dissatisfaction with how choropleth maps were used to understand economic data from the BLS. Linked micromaps provide visual summaries of statistics, data, and geographic distributions in a coherent and intuitive manner. They enable the exploration and dissemination of statistical information (e.g., complex distributional and multivariate patterns) connected with geography in an intuitive way. In essence, linked micromaps consist of several graphical columns, with one column depicting a set of small maps of areal or polygonal units, with one unit represented by each row. These units are linked across rows using color to other columns showing statistics of interest. In addition, the rows are sorted by a specified variable (e.g., a measure of size) associated with a given unit. Different sub-region shading options are available in the `micromapST` package for exploring cumulative spatial distributions. The resulting plots are intended to provide an intuitively accessible overview of:

(a) Geographical patterns and spatial association, e.g., ways in which the statistics of interest may be similar or different across nearby units.
(b) Ordinal patterns, i.e., ways in which the reported statistics may vary with respect to the variable used for the sort order.

*Elements of Linked Micromaps.* Linked micromaps consist of at least three vertical columns containing (1) a series of small maps within the main geographical area (e.g., the US, single state, Africa, etc.) with colored or highlighted regions indicating the current row (e.g., a state), (2) a legend naming each areal unit, and (3) one or more columns displaying a statistic of interest (e.g., number of establishments for a given industry). Figure 1.1 presents an example of linked micromaps showing statistics such as the total number of establishments in the first quarter of 2025 and one-year percent changes in wages and employment for all industries in the U.S. We order the states by the number of establishments to give a sense of the size of the underlying population, noting that other sort variables can be used to explore the data, search for structure, and to suggest relationships. The color of the regions in the small maps links the statistic along the rows and columns to the region indicated in the micromaps. The color does not correspond to a quantitative measure; it serves as an index linking regions to data.

[**FIGURE 1.1**]

**Figure 1.1** This is an example of a linked micromaps graphic produced by the `micromapST` package using data from QCEW for all industries from Q1 2020 through Q1 2025. These plots enable a data practitioner to explore trends and relationships among the number of establishments, employment, and wages. The leftmost data column reflects the skewed distribution of state-level establishment counts and regional patterns. The second data column displays an interesting sawtooth pattern in the time series plots, which is a well-known sawtooth pattern for wage data. The third data column plots the one-year (2025 Q1 vs. 2024 Q1) percent change in wages against the percent change in employment counts. The scatterplots display three variables, which are the establishment counts via the filled circles, wages, and employment. This demonstrates how linked micromaps can help one compare multivariate relationships across geography. Figure 1.1 also gives examples of possible plot types available in the `micromapST` package, which will be explored in later sections.

The small sub-regions in the micromaps are organized into perceptual groups to make it easier to understand the graphic and to extract useful information, as we will see in later sections. It is important to remember that the sort variable determines which regions are grouped together. The US regions are readily divided in a rather pleasing and intuitive way. First, they are divided according to the median value of the ordered statistic, with 25 regions above and 25 below. Those are further subdivided into groups of five. Note that the region corresponding to the median sorted variable is in its own group; see Figure 1.1. Showing just five states in each group helps the viewer to identify the states, especially the small ones like Rhode Island or Washington, DC that are easier to miss in a single choropleth map of all states. Figure 1.2 provides an enlarged view of the first perceptual group from Figure 1.1 with the elements of linked micromaps labeled.

[**FIGURE 1.2**]

**Figure 1.2** This shows an enlarged view of the first perceptual group of Figure 1.1, with labels indicating the elements of linked micromaps.

While original tools for linked micromaps used Java (Carr and Pickle, 2010), statisticians and data scientists can easily create them now using two R packages – `micromap` and `micromapST`– which are available on CRAN (Payton, et al., 2015; Pickle, et al., 2015). Pickle and Pearson (2015) developed the `micromapST` package as an easy-to-use interface for creating linked micromaps based on a specified geographical classification. The `micromapST` package provides options for the following types of graphics: arrows, bars, boxplots, dots, segmented bars for categorical data, scatterplots, and time series. More information on the package, available plot types, glyphs, examples, and other options (e.g., conveying uncertainty) can be found in Pickle, et al. (2015).

The current article presents examples where the geographies of interest are the 50 states and Washington, DC in the United States. The `micromapST` package was developed with these geographies in mind and was used to create the graphics in this article. R scripts to create the linked micromaps included herein, along with the publicly available data will be available on github https://github.com/wendylmartinez/Linked-Micromaps-JSTP26/.

Color is an important part of linked micromaps, as it serves as a link between the geographic regions and the statistics. The graphics in the print version of this article will be in grayscale. Color images of the figures are available in the online version of the article and on the github site. Note that the online graphics use a palette that is safe for viewers with color vision issues, which is not the default palette in the `micromapST` package; see the script files for more details.

## 2. Comparison of Subpopulations

There is often a high degree of heterogeneity of economic and social conditions within high-level geographies. Consequently, graphics or tables that just compare whole states (or other large geographical areas) may not tell the entire story about industrial or geospatial structure.

This section explores this issue in three ways: comparison of industrial sectors across US states and across time; comparison of wages in different sub-geographies; and comparison of employment and wages over individual counties in New York state.

## 2.1 Comparing Multiple Industrial Sectors

The following example of a linked micromaps graphic uses data from the QCEW during the COVID-19 pandemic era. These data raise questions about the effect of this pandemic on employment. How did reported employment change during the early years of the pandemic? Did those changes show different patterns across different industries? Were the patterns of change similar across states? To explore these issues, we examined data from two industries known to have volatile employment patterns during the pandemic period: Leisure and Hospitality (QCEW super-sector 1026) and Construction (QCEW super-sector 1012).

Powers, et al., (2024) provides an example where four choropleth maps are obtained using the QCEW tool to explore the over-the-year percent change in employment and discusses the limitations of such an approach. The most important limitation of such a sequence of choropleth maps is the inability to see the actual state employment values over time and on a consistent visual basis, since the legend changes with each quarter. We can easily visualize employment over time and geography (space) using linked micromaps.

In Figure 2.1, the states have been sorted based on the 2021 Q2 Total Employment in all industries. The first data column presents a time plot of the quarterly values of over-the-year change in employment in the Leisure and Hospitality industry from the first quarter in 2020 to the fourth quarter of 2021. The second data column shows the corresponding time plot for the Construction industry.

The linked micromaps provide information on the individual state values, and we can see magnitudes of the change over time. The plots indicate that the downturn in Leisure and Hospitality jobs was almost immediate across the US, following the broad public awareness of the COVID-19 pandemic and related lockdowns. In contrast, within the Construction industry, the changes in employment during the early pandemic quarters displayed substantial variability across states. Of special note is the pattern of heterogeneity displayed among the five perceptual groups at the bottom of the plot (especially for Construction), i.e., the 25 states with the lowest total employment that might warrant further investigation by subject matter experts.

[**FIGURE 2.1**]

**Figure 2.1** This figure is a linked micromaps plot comparing the percent over-the-year change in employment during the first two COVID-19 years for two industries. The states are ordered by the Total Employment for All Industries in 2021 Q2. Shocks are visible when the quarantine and recovery started and indicate some stabilization in subsequent months. While the two industries have different vertical axis scales, the departure from zero can be easily compared. In addition, the time plots for the two industries display notably different patterns of across-

state heterogeneity that might warrant further exploration by subject matter experts. For example, Hawaii shows an increase in mid-2021 for Leisure and Hospitality, but not for Construction. In addition, Construction exhibits heterogeneity across states in its temporal patterns of employment change, especially among the states with smaller total employment (i.e., lowest perceptual groups).

## 2.2 Comparing Wages over Multiple Sub-Regions

Figure 2.2 presents information from the OEWS program for the Police and Sheriff Patrol Officers occupation (Standard Occupational Classification code 33-3051). States are sorted by the estimated mean hourly wage for the state. Wages for this occupation generally are not expected to be homogeneous within a given state, especially when comparing police officers working in rural versus urban areas. To explore this, the second data column presents the range of mean wages for each Metropolitan Statistical Area (MSA) within a given state. Different states have different degrees of across-MSA variability in their estimated mean wages; patterns of variability within California and Florida are especially notable.

Similarly, the third data column shows the range of mean wages for Balance of State (BOS) areas within a state. A given state may have several distinct BOS areas, which tend to be rural or semi-rural. It is of interest to explore the extent to which BOS areas have wage rates that are relatively low compared with MSAs in the same state. States also display different degrees of across-BOS variability in their estimated mean wages, although the differences are less pronounced than for the across-MSA pattern.

The dashed vertical line in the first three data columns indicates the national mean wage for Police and Sheriff Officers. This enables the comparison of state, MSA, and BOS wages with the national average. Those states with lower wages tend to be in the South (see the bottom two perceptual groups), and wages in MSAs tend to be closer to the national average wage as compared to BOS wages (see the middle two data columns).

The final column of Figure 2.2 displays a scatterplot of the mean wages (arithmetic average of the MSA-specific and BOS-specific mean wages over all MSAs and all BOS areas) for MSA (vertical axis) and BOS (horizontal axis). For each perceptual group, the applicable plotted points are displayed with the colors of the specified states, while the other plotted points are black-and-white open circles. The plot displays a general pattern of positive association between the MSA and BOS estimated mean wages in each state.

[**FIGURE 2.2**]

**Figure 2.2.** This linked micromaps plot compares wages for Police and Sheriff's Patrol Officers (Standard Occupational Classification: 33-3051) at the state level and for sub-regions at the MSA (urban) and BOS (rural) levels. States are ordered by their average hourly wage for the state and shown as dots in the first data column. The next two columns display the range of the hourly wage for the urban and rural areas as arrows. The vertical dashed line shows the national hourly wage for this occupational classification. It would be of interest to explore the extent to which the differing degrees of heterogeneity across states in the MSA and BOS mean wage rates may be attributable to specific factors like collective bargaining legislation and competitive market forces. The fourth data column is a scatterplot comparing the wages for urban areas (MSA) versus rural areas (BOS). The plot displays a general pattern of positive association between the MSA and BOS estimated mean wages in each state.

## 2.3 Comparing Geographical Units Other Than States

Most of the linked micromaps in this paper focus on comparisons across the 50 states and the District of Columbia in the United States. However, the latest version of the `micromapST` package can be used for any geographies with user-supplied shape files. Examples include

- Provinces, and sub-provincial areas in Canada, based on boundary and shape-file information like that provided by Statistics Canada at https://www12.statcan.gc.ca/census-recensement/2021/geo/index-eng.cfm.

- Nations and sub-national regions within the European Union, based on boundary and shape-file information found here https://ec.europa.eu/eurostat/web/gisco/geodata.

- U.S. geographic areas and boundaries for counties, tracts, etc. found at the Census Bureau website https://www.census.gov/geographies/mapping-files/time-series/geo/cartographic-boundary.html.

To illustrate this, Figure 2.3 presents a linked micromaps graphic for the 62 counties within New York state. Counties are sorted by their estimated total population in July 2024, as given here: https://www.census.gov/data/tables/time-series/demo/popest/2020s-counties-total.html. Each perceptual subgroup includes four or five counties.

The two data columns display data from the QCEW manufacturing sector (QCEW super-sector 1013). These data were downloaded from the QCEW website using the tool discussed in the first section. These data columns show that most counties had declines in employment, along with increases in mean wages for the manufacturing sector. These changes do not appear to have a strong association with the sort variable (overall population size of the counties). It is important to note that the sort variable is not explicitly displayed in a data column, as we have in other examples in this article. Thus, the sort variable in this case provides information about another statistic not explicitly shown, which is the rank order of the counties based on population size.

[**FIGURE 2.3**]

**Figure 2.3** This illustrates a linked micromaps plot for a different geography using the `micromapST` package. The package comes with some alternative geographies and accommodates user-supplied boundary files. Here we display the one-year changes in employment and wages in Manufacturing for counties in New York state for Quarter 4 2024 (i.e., 2024 Q4 vs. 2023 Q 4) based on the Quarterly Census of Employment and Wages (QCEW). The counties are sorted by their 2024 population count, the value of which is not shown in any of the data columns. Wages increased in most counties, while employment decreased. Although, this relationship did not hold for all. It would be of interest to explore the extent to which these different patterns may be attributable to changes in specific sub-industries within manufacturing. Finally, note the extreme decrease in employment and wages for Hamilton County, which is the smallest with a population of around 5,000.

## 3. Visualization of Multivariate Data

Section 2 considered the use of linked micromaps to explore the behavior of one variable over multiple subpopulations, such as multiple industries and multiple geographical areas. One may also use linked micromaps to explore multivariate relationships. Of special interest is exploring the extent to which such relationships may differ within and across perceptual groups. This section illustrates how this can be done in linked micromaps through scatterplots, boxplots, sort order, and more.

### 3.1 Nonparametric Regression with a Scatterplot Smooth

Analysts often need to evaluate the extent to which an industry or occupation may be concentrated in a geographical area and to explore the extent to which wages or other conditions may be associated with it. One relatively simple measure for understanding how occupations are concentrated in sub-regions is the “location quotient” (LQ) as discussed in https://www.bls.gov/cew/about-data/location-quotients-explained.htm. For a given geographical area A and industry-occupation category B, one defines the ratios,

*Local concentration = (Total employment in category B within area A)/(Total employment within area A)*

*National concentration = (Total employment in category B within entire country)/(Total employment within entire country)*

*and*

*Location quotient (LQ) = (Local concentration)/(National concentration)*

Thus, for the specified area A and category B, a location quotient greater than 1 will indicate that area A has employment in category B that is disproportionately high, relative to employment in category B nationwide. In other words, a location quotient greater than 1 is a ‘hot spot’ for that category compared to the national level.

[**FIGURE 3.1**]

**Figure 3.1** We return to the Police and Sheriff Patrol Officers data in this linked micromaps graphic where three of the columns for hourly wages at the state, MSA, and balance of state (BOS) levels are repeated. These columns provide context for the bivariate relationship displayed in the first data column, which shows a scatterplot of the location quotient versus the state mean hourly wage for this occupation. A lowess smooth is added, which suggests a nonlinear relationship between employment levels and wages. The elbow on the curve appears to be close to the national mean hourly wage. One could explore further the extent to which the observed relationship between location quotient and mean hourly wage (first data column) may be attributable to the factors such as collective bargaining legislation in the state or competitive market forces.

Figure 3.1 includes the wage information for the Police and Sheriff Patrol Officers we saw in Figure 2.2 with an additional column showing a scatterplot of the state-level location quotient for the Police and Sheriff Patrol Officers occupation (vertical axis) against the state-level mean hourly wage rate for that occupation. The scatterplot suggests a nonlinear relationship between the state LQs and wages. The `micromapST` package has the functionality to add a local linear lowess (locally weighted scatterplot smoothing) regression curve fit to the scatterplot. A lowess scatterplot smooth is a nonparametric regression approach that uses local linear fits to construct the curve through the points (Cleveland, 1979).

To explore this relationship further, Figure 3.2 shows the scatterplots in two graphs. The top panel repeats the scatterplot from the first perceptual group in the linked micromaps graphic in Figure 3.1. The bottom panel presents the same underlying scatterplot but shows two lowess smooths with different smoother spans. A larger span gives a smoother curve, while smooths from smaller spans usually have more wiggles from fitting the data more closely. It is always a good idea to vary parameters like the span when exploring data to see what insights can be gained.

This example shows how linked micromaps can provide a broad suite of tools for the initial exploration of patterns of variability in multiple dimensions before further analysis or model building is done. When the linked micromaps help us to identify especially interesting patterns of variability, we can use further tools (e.g., lowess plots with other spans, R `ggplot2` smooths with confidence bands) to explore those patterns in additional detail.

[**FIGURE 3.2**]

**Figure 3.2** This figure shows two scatterplot smooths based on the Police and Sheriff Patrol Officers data. The graph at the top is the scatterplot from the highest perceptual group in the linked micromaps plot in Figure 3.1. This includes a lowess curve created using a default smoother span of 2/3. The bottom graph explores the extent to which the curve may be sensitive to the choice of span. Like the one from `micromapST`, the bottom graph uses the lowess function (local linear fit) in base R. The black solid curve is the same one in the top graph, with a default span of 2/3. The dashed red line uses a span of 0.3. On the far-left side of the graph, the dashed red line is somewhat higher than the solid black line, reflecting the effects of some points that have relatively high LQ values. In addition, the dashed red line displays somewhat greater variability in the middle of the graph for units relatively close to the mean hourly wage (vertical green line). It is important to explore data sets by varying models and parameters to gain insights into the data and potential relationships to propose.

## 3.2 Showing Three Variables in a Scatterplot

This next example also explores wage and job data from the OEWS, this time looking at elementary school teachers. Figure 3.3 shows a linked micromaps graphic displaying state-level information for educators in two complementary categories, described as General Education Teachers (Elementary School Teachers, Excluding Special Education, SOC: 25-2021) and Special Education Teachers" (SOC: 25-2056), respectively. The states are sorted by their location quotients for special education teachers. There is a roughly symmetric distribution of the LQs around the median.

There do not appear to be any strong patterns of concentration of exceptionally high or low LQs for special education teachers across the regions of the United States.

Two data columns display scatterplots exploring relationships of the wages for General and Special Education Teachers. The first scatterplot, which is in the middle data column, compares the LQs for General Education Teachers (vertical axis) against LQs for Special Education Teachers (horizontal axis). Just looking at the scatterplot (ignoring the curve), one would think there is not much of a relationship between employment levels for the two categories. The points look like a horizontal band. However, the lowess smooth points to a possible nonlinear relationship with mild curvature, which may warrant further exploration. Readers are encouraged to create curves using different smoother spans or other nonparametric regression methods (e.g., loess), as we mentioned with the Police and Sheriff data in Figure 3.2. The R code showing how to do that can be found in the script files (see https://github.com/wendylmartinez/Linked-Micromaps-JSTP26/) for Figure 3.3.

The last column shows scatterplots of estimated state-level mean wages for General

Education Teachers against those for Special Education Teachers. There is a clear linear relationship with a positive slope. By default, the line $y = x$ is shown in the background, and we see that the data fall approximately along that line. So, a lowess smooth seems unnecessary in this case.

The scatterplots show the relationship between wages for the different types of teachers. Elementary teachers appear to have similar mean wages in a state regardless of the type of students being taught. However, there is a third variable shown in these scatterplots, which is represented by the filled in circles. These are indexed to each state's LQ (employment) for special education elementary school teachers, not the wages, which are represented on the two axes. This helps us explore potential and interesting relationships between wages (the two axes) and employment (LQs as the sort variable).

For example, West Virginia has a location quotient for special education that is the highest among the states. However, wages for elementary school teachers in West Virginia are among the lowest in the US, as indicated by the highlighted dot in the lower left corner of the scatterplot. Conversely, Washington state and Oregon are among the high-paying states (in the top 4) for special education teachers and have a relatively low concentration of special education teachers.

Economists and policy makers can use the information in the linked micromaps to locate interesting relationships for improved decision making. Relationships like these would be difficult if not impossible to find if the data were shown in other ways like tables or bar charts.

[**FIGURE 3.3**]

**Figure 3.3** This linked micromaps plot displays the location quotients (LQs) and wages for elementary school teachers of general education classes and special education. The states are sorted based on the LQ for special education teachers as shown in the first data column. Consequently, the states in each perceptual group will have similar location quotients. The next column contains scatterplots of the LQ for general education teachers versus the LQ for special education teachers. Adding a lowess smooth hints at a slight nonlinear relationship. Subject matter experts might explore deviations from this lowess curve. For instance, in the last perceptual group, MO and AL are above the line and might be potential outliers. It would be of interest to explore the extent to which the deviations might be attributable to regional differences within the state. The wages for the two categories of teachers are compared in the scatterplots displayed in the last column and are showing an approximate linear relationship. These scatterplots display three variables – wages (via the two axes) and employment of special education teachers (via the filled in circles). One interesting state is WV, which has the highest LQ for special education teachers, but whose wages are at the lowest end of the range. As with Figure 3.1, readers are encouraged to try different smoother spans with the lowess. See the files on the github site for example code.

### 3.3 Expanded Visualization of Tabular Information: Multiple Indicators of Variability

Table 1 OEWS 2023 Employment and Hourly Wages for Software Developers (Source: https://www.bls.gov/oes/2023/may/oes151252.htm.)

| STATE | LOCATION QUOTIENT | H_MEAN | MEAN_PRSE | H_PCT10 | H_PCT25 | H_MEDIAN | H_PCT75 | H_PCT90 |
|---|---|---|---|---|---|---|---|---|
| AL | 0.76 | 53.19 | 1.5 | 29.58 | 37.73 | 49.39 | 64.57 | 81.29 |
| AK | 0.08 | 70.01 | 4.6 | 41.95 | 52.27 | 72.79 | 84.83 | 93.45 |
| AZ | 1.13 | 61.56 | 1.6 | 37.72 | 45.77 | 59.22 | 69.85 | 84.83 |
| AR | 0.41 | 42.37 | 4.8 | 14.13 | 27.96 | 44.3 | 53.97 | 64.05 |
| CA | 1.55 | 83.55 | 0.8 | 49.64 | 65.04 | 81.09 | 100.92 | 108.97 |
| CO | 1.47 | 69.92 | 1.1 | 41.62 | 50.73 | 64.89 | 80.44 | 99.25 |
| CT | 1.01 | 61.75 | 1.5 | 37.51 | 47.93 | 60.14 | 75.69 | 89.63 |
| DE | 0.95 | 63.29 | 2 | 44.08 | 51.97 | 63.31 | 73.29 | 84.04 |

Data providers typically disseminate data in tables that can have many variables. It is difficult to tease out important information even when a partial set of variables (or columns) are explored as seen in Table 1, where we display a subset of US states. We will show next how all the variables in Table 1 and more can be displayed in a linked micromaps plot.

Figure 3.4 presents information on employment and hourly wages for the Software Developers occupation (SOC: 15-1252) based on the OEWS 2023 survey. States are sorted by their location quotients for this occupation, which are displayed as dots in the first data column. A vertical green line at 1 is provided as a reference value. Recalling that an LQ greater than 1 indicates that the state has a higher concentration of jobs in that occupation as compared to the nation, while an LQ less than 1 indicates a lower concentration.

The next two variables from Table 1 to be shown in the linked micromaps are the hourly mean wage (H_MEAN) and the standard error (MEAN_PRSE). These are presented in the second data column of Figure 3.4 as dots representing the state-level mean hourly wage and bars showing 90% confidence intervals. The national mean hourly wage of $66 is shown as a green vertical line and can be used as a reference value.

The remaining columns of Table 1 contain five percentiles that we show as elements of a boxplot in the last column of the linked micromaps in Figure 3.4. The edges and middle lines of the boxes are the usual quartiles (H_PCT25, H_MEDIAN, H_PCT75). The ends of the whiskers are the 10% and 90% percentiles (H_PCT10 and H_PCT_90). This data column in the linked micromaps shows a vertical line for the reference value corresponding to the national *median* wage.

The linked micromaps graphic shown in Figure 3.4 enables state-level comparisons of the employment location quotients with both the mean structure and the percentile structure for hourly wages. Finally, note that Figure 3.4 displays simultaneously several complementary sources of information regarding variability, including variability in the employment location quotients and wage patterns across states; uncertainty in the mean estimates, as reflected in their 90% confidence bands; and within-state variability of the wage percentiles, as indicated in the final column.

[**FIGURE 3.4**]

**Figure 3.4** This figure shows how linked micromaps can be used to display many types of variability in an intuitive way. This graphic shows employment (sort variable) and wages for the Software Developers occupation. Data from all nine columns in Table 1 are visualized in this linked micromaps plot. Additional reference values relevant to each column are shown as vertical green lines to help assess distribution information such as location and variability as compared to national values. Note especially that the middle data column displays state-level mean wage estimates, accompanied by 90% confidence intervals, and thus conveys visually one important measure regarding variability. In addition, the rightmost column presents state-level percentiles for the wage data, thus conveying a second, complementary indication of variability in the available wage information. Including both data columns in the linked micromaps enables the exploration of both across-state patterns of mean wages and of heterogeneity of wages within states.

## 4. Discussion

This paper has noted several ways in which linked micromaps can help one explore features of official statistics related to geographical areas. These included relatively simple descriptive statistics, comparison of subpopulations (Section 2), and multivariate analyses (Section 3). These ideas and options for data visualization are applicable to many potential workflows of data practitioners. Of special interest was the use of these approaches to explore several complementary notions of variability, including variability across different geographical areas, occupations, industries and outcome variables; and variability represented by standard errors and quantiles. As with many tools for exploratory data analysis, linked micromaps can be used to identify interesting data features and to suggest relationships that warrant further study through more formal statistical inference and inputs from domain experts. There are numerous opportunities for further development and use of linked micromaps in statistical practice as presented in the following sub-sections.

### 4.1. Choice of Sort Order: An Additional Dimension of Multivariate Exploration

With the examples of linked micromaps shown in sections 1 through 3, states or counties were sorted according to relatively simple measures of size or functions thereof. In general, the sort order helps us to understand the extent to which the variables (and associated multivariate relationships) displayed in each column are associated with the sort variable. The sort order also determines the membership in perceptual subgroups and thus focuses our attention on variability (or similarity) across states that have similar values of the sort variable. For general background on sort order in statistical graphics, see, e.g., Robbins (2013, pp. 160-161). In addition, see Carr and Pickle (2010, pp. 2-3) for some specific comments on sort orders for micromaps.

For some linked micromaps, there may be several potential candidates for a sort variable, such as an overall unit size, absolute or relative change in that size, other measures of underlying population dynamics, an aggregate response rate, or another measure of information quality (e.g., timeliness). In addition, for cases that involve parametric or nonparametric regression analyses (see Section 3), one may consider using the state-level regression residual as a sort variable. This can help identify additional regressors to include in expanded versions of the model. Exploratory comparison of the linked micromaps based on these alternative sort variables can also offer further insights for subject matter experts.

Finally, for cases that involve a substantial number of candidate sort variables, one can carry out a principal-component analysis of those candidate variables, and then examine linked micromaps that use, respectively, sorting on the first, second, or third principal components, for example. Due to the orthogonality of the principal components, these different sort orders may provide a relatively efficient way to screen for prospective relationships between sort variables and the features displayed in subsequent columns of the linked micromaps display.

### 4.2. Geographical Units with Irregular Shapes or Highly Heterogeneous Sizes

In some cases, geographical units may vary greatly in area or may have very irregular shapes, and those features may lead to complications in the use of linked micromaps. Consequently, in some cases, one might consider modifying the topology or size of the regions displayed in the micromaps in ways that could improve the legibility and interpretability of irregularly sized or shaped geographical areas. For instance, one could replace a very irregularly shaped sub-region with a more compact simple shape, with appropriate cautionary notes for the reader. This approach has several potential strengths and limitations. For example, previous work with modification of maps to address irregular unit sizes has developed within the literature on cartograms. For some discussion of that literature and related critiques see Tobler (2004), Nusrat et al. (2018 a, b), Duncan and Gastner (2024) and references cited therein.

4.3. Non-Hierarchical Geographical Structures.

For many social and economic phenomena, some important geographical patterns may not fit with strict hierarchies. For example, many U.S. statistical agencies publish estimates for individual states, individual counties within states, and metropolitan areas. Metropolitan Statistical Areas can involve multiple counties across multiple states. For instance, the Chicago-Naperville metropolitan statistical area includes portions of the states of Illinois, Indiana and Wisconsin. In addition, the boundaries of metropolitan areas often change over time. Those changes lead to important questions about interpretable ways to account for those changes in time plots embedded within linked micromaps. Related comments apply to “off spine” geographical areas that are important in analysis of disclosure protection methods for hierarchical geographical structures, as described in Cumings-Menon et al. (2025). Refinement of linked micromaps tools to address each of these issues can expand the scope and impact of these graphical methods.

4.4. Applications to Multivariate Conceptual Spaces.

In addition, one could apply the general ideas of linked micromaps to the graphical analysis of objects or zones in multivariate conceptual spaces, rather than geographical areas. A simple example would involve cells defined by the intersection of two ordinal variables, e.g., age groups and educational attainment groups. In the resulting linked micromaps, those cells would replace the states, and other variables (e.g., income, employment rates or health conditions) would be presented for each cell. As with previous examples involving states, there could be special interest in visual exploration of potential adjacency phenomena, as well as general gradient patterns across the ordinal variables that may define the cells.

4.5. Application to Model Building, and Exploration of Additional Methodological Features

One may use linked micromaps in model building and to explore methodological features involved in the production of official statistics. For instance, in an extension of the ideas developed in this paper, one may consider side-by-side display of multiple measures of variability including standard errors of mean estimates, quantiles of underlying unit measures, and heterogeneity of subpopulation means. In addition, one may use linked micromaps to compare competing methods of weighting and variance estimation. For example, one may use variants on linked micromaps to explore strata and adjustment cells used for sample design and estimation. These strata and cells may involve geographical groups, plus conceptual groups (e.g., demographics, industry, and occupation). These groupings often involve collapsing procedures that prevent sparse cells, and linked micromaps can help one explore prospective options. Linked micromaps can also support the exploration of other data quality measures as they relate to geographical units. Examples of such measures include diagnostics related to informative sampling, nonresponse bias and other incomplete-data issues (Little and Rubin, 2019), non-probability samples (Pfeffermann and Sverchkov, 2025), and "total survey error" models (Amaya et al., 2020).

### 4.6. Exploration of Trade-Offs Among Multiple Dimensions of Quality, Risk and Cost.

Linked micromaps can enrich the set of tools that statistical agencies use to explore practical trade-offs in their survey methodology decisions. For instance, one can display multiple dimensions of cost and quality that are important for decisions on the fine-tuning of adaptive survey design or imputation. Another example involves the display of risk and quality measures that inform decisions on privacy budget allocation for disclosure protection.

### 4.7. User-Friendly Implementation via R Shiny

The linked micromaps approach was originally developed as an improvement to choropleth maps for displaying statistical summaries connected with spatial areal units, such as countries, states, and counties. As shown in this article, the `micromapST` package provides useful ways for data analysts and statistical practitioners to explore their data before further analysis. The `micromapST` package takes care of most details of the layout, but it requires a significant initial investment in learning how to specify the data frames to create the desired graphic, as can be seen in the script files provided with this article (see https://github.com/wendylmartinez/Linked-Micromaps-JSTP26/).

Exploring data is fundamentally an iterative process, where the practitioner should try different orderings, plot types, statistics, etc. in their search for structure and insights. A graphical user interface (GUI) can make exploratory visualizations easier, faster, and more intuitive. This is the motivation behind the R Shiny `micromapST` app, which is expected to be published in 2026. See Powers and Martinez (2025) for a tutorial introduction to the app.

**Acknowledgments and Disclaimers:**

The authors thank reviewers from the Bureau of Labor Statistics and the Census Bureau for helpful comments. In addition, the authors thank an anonymous referee for several literature references, a suggestion to add discussion of the smoother span in Figure 3.2, and other very helpful suggestions. The views expressed in this paper are those of the authors and do not reflect the policies of the U.S. Census Bureau nor of the U.S. Bureau of Labor Statistics.

All data and code used for this paper are in the public domain and are available through the links provided in Section 1 and on the github site.

**Figure 1.1** This is an example of a linked micromaps graphic produced by the `micromapST` package using data from QCEW for all industries from Q1 2020 through Q1 2025. These plots enable a data practitioner to explore trends and relationships among the number of establishments, employment, and wages. The leftmost data column reflects the skewed distribution of state-level establishment counts and regional patterns. The second data column displays an interesting sawtooth pattern in the time series plots, which is a well-known sawtooth pattern for wage data. The third data column plots the one-year (2025 Q1 vs. 2024 Q1) percent change in wages against the percent change in employment counts. The scatterplots display three variables, which are the establishment counts via the filled circles, wages, and employment. This demonstrates how linked micromaps can help one compare multivariate relationships across geography. Figure 1.1 also gives examples of possible plot types available in the `micromapST` package, which will be explored in later sections.
RETURN

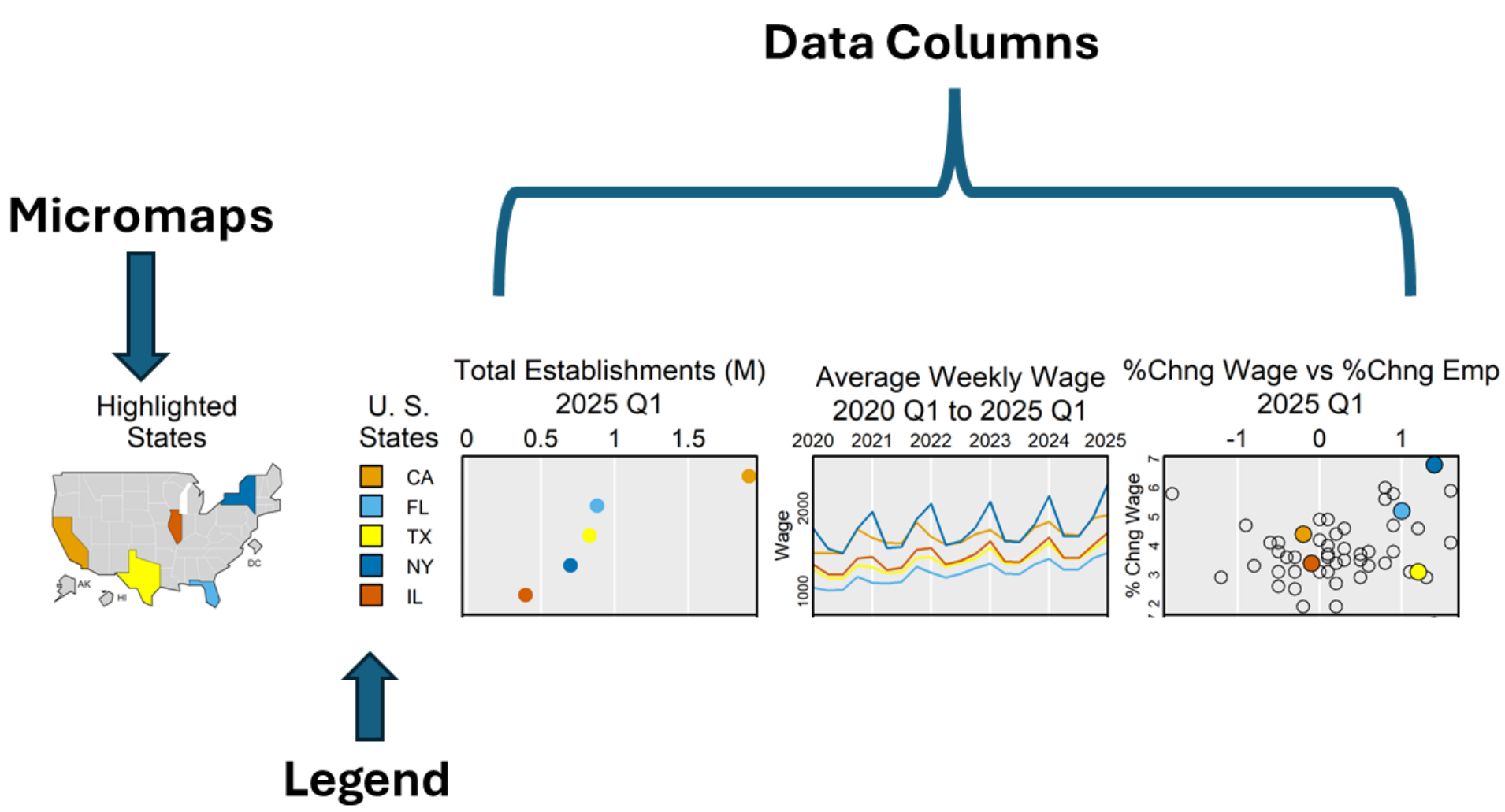

Figure 1.2 This shows an enlarged view of the first perceptual group of Figure 1.1, with labels indicating the elements of linked micromaps. RETURN

QCEW Over-the-Year Change in Employment
Sorted by Total Employment All Industries, 2021 Q2

**Figure 2.1** This figure is a linked micromaps plot comparing the percent over-the-year change in employment during the first two COVID-19 years for two industries. The states are ordered by the Total Employment for All Industries in 2021 Q2. Shocks are visible when the quarantine and recovery started and indicate some stabilization in subsequent months. While the two industries have different vertical axis scales, the departure from zero can be easily compared. In addition, the time plots for the two industries display notably different patterns of across-state heterogeneity that might warrant further exploration by subject matter experts. For example, Hawaii shows an increase in mid-2021 for Leisure and Hospitality, but not for Construction. In addition, Construction exhibits heterogeneity across states in its temporal patterns of employment change, especially among the states with smaller total employment (i.e., lowest perceptual groups). RETURN

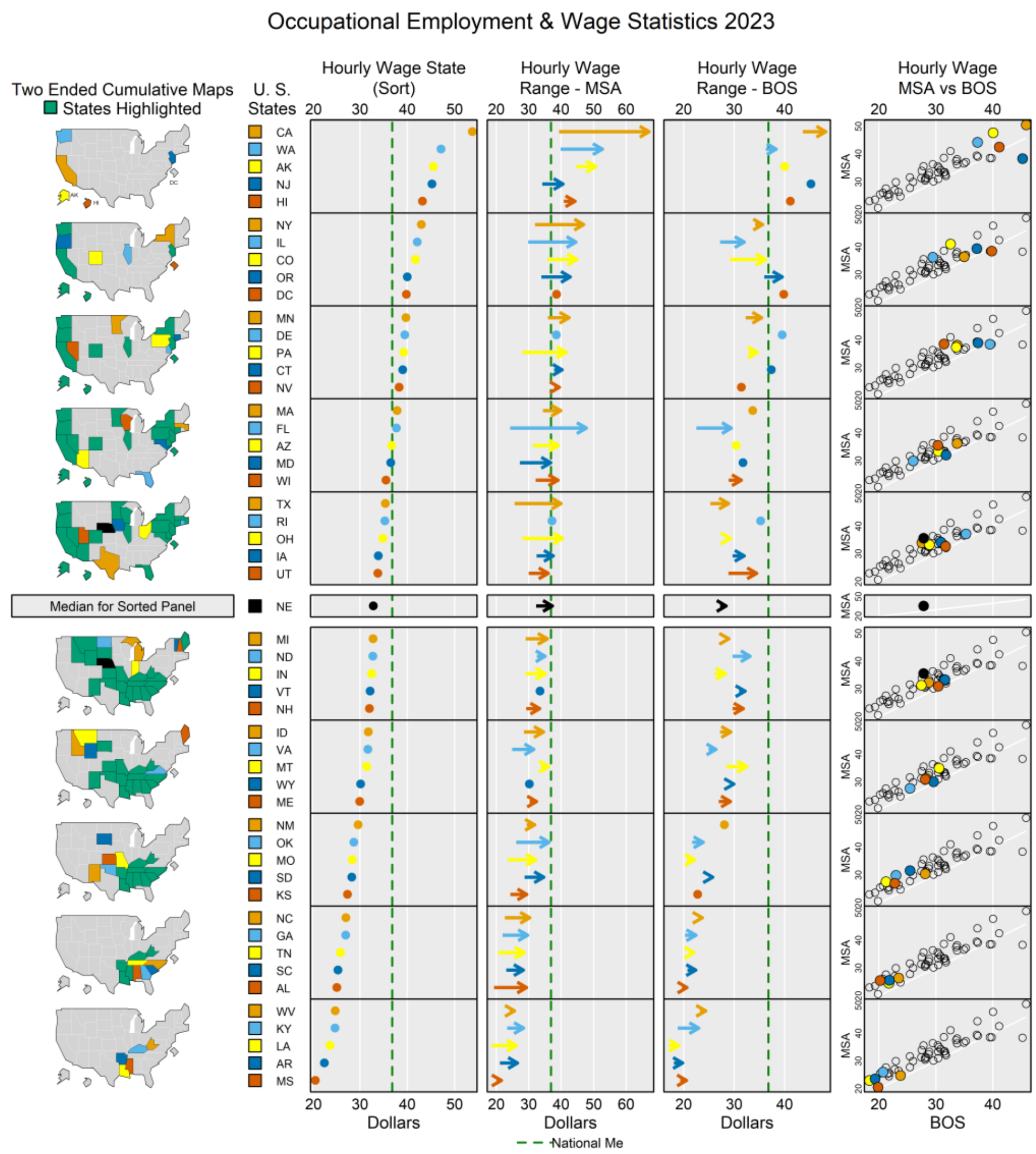

**Figure 2.2.** This linked micromaps plot compares wages for Police and Sheriff's Patrol Officers (Standard Occupational Classification: 33-3051) at the state level and for sub-regions at the MSA (urban) and BOS (rural) levels. States are ordered by their average hourly wage for the state and shown as dots in the first data column. The next two columns display the range of the hourly wage for the urban and rural areas as arrows. The vertical dashed line shows the national hourly wage for this occupational classification. It would be of interest to explore the extent to which the differing degrees of heterogeneity across states in the MSA and BOS mean wage rates may be attributable to specific factors like collective bargaining legislation and competitive market forces. The fourth data column is a scatterplot comparing the wages for urban areas (MSA) versus rural areas (BOS). The plot displays a general pattern of positive association between the MSA and BOS estimated mean wages in each state. RETURN

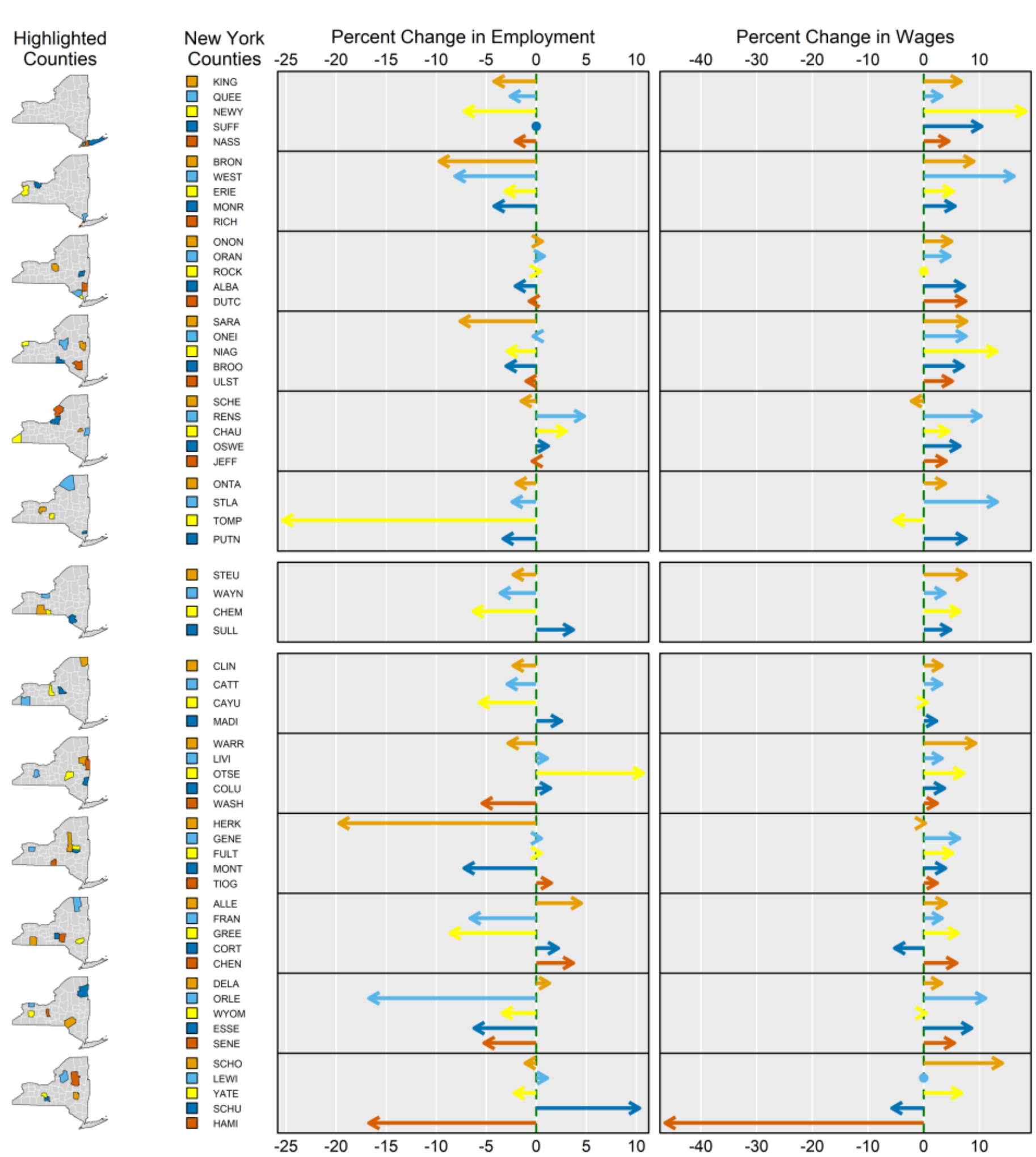

**Figure 2.3** This illustrates a linked micromaps plot for a different geography using the `micromapST` package. The package comes with some alternative geographies and accommodates user-supplied boundary files. Here we display the one-year changes in employment and wages in Manufacturing for counties in New York state for Quarter 4 2024 (i.e., 2024 Q4 vs. 2023 Q 4) based on the Quarterly Census of Employment and Wages (QCEW). The counties are sorted by their 2024 population count, the value of which is not shown in any of the data columns. Wages increased in most counties, while employment decreased. Although, this relationship did not hold for all. It would be of interest to explore the extent to which these different patterns may be attributable to changes in specific sub-industries within manufacturing. Finally, note the extreme decrease in employment and wages for Hamilton County, which is the smallest with a population of around 5,000. RETURN

**Figure 3.1** We return to the Police and Sheriff Patrol Officers data in this linked micromaps graphic where three of the columns for hourly wages at the state, MSA, and balance of state (BOS) levels are repeated. These columns provide context for the bivariate relationship displayed in the first data column, which shows a scatterplot of the location quotient versus the state mean hourly wage for this occupation. A lowess smooth is added, which suggests a nonlinear relationship between employment levels and wages. The elbow on the curve appears to be close to the national mean hourly wage. One could explore further the extent to which the observed relationship between location quotient and mean hourly wage (first data column) may be attributable to the factors such as collective bargaining legislation in the state or competitive market forces. RETURN

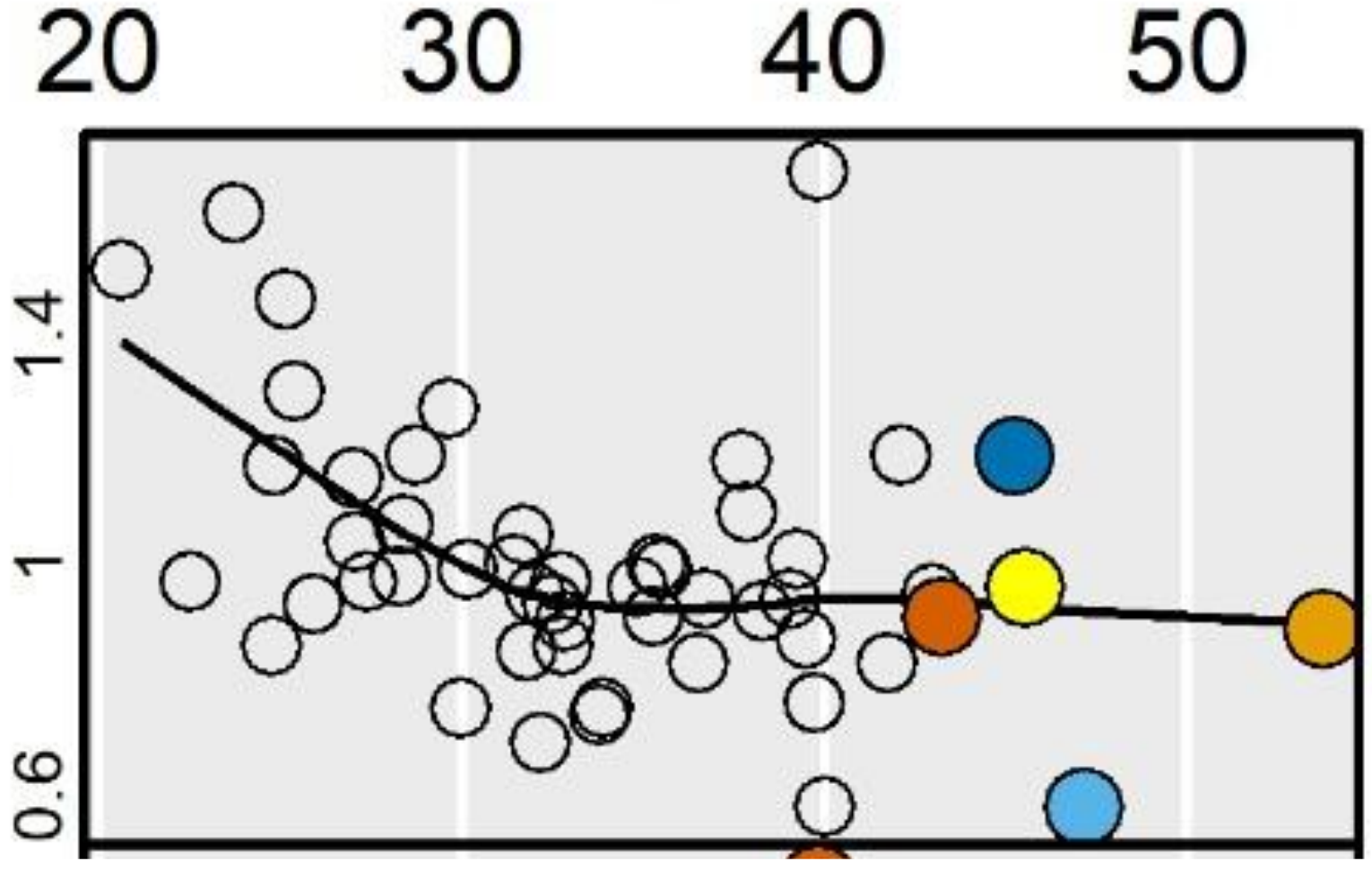

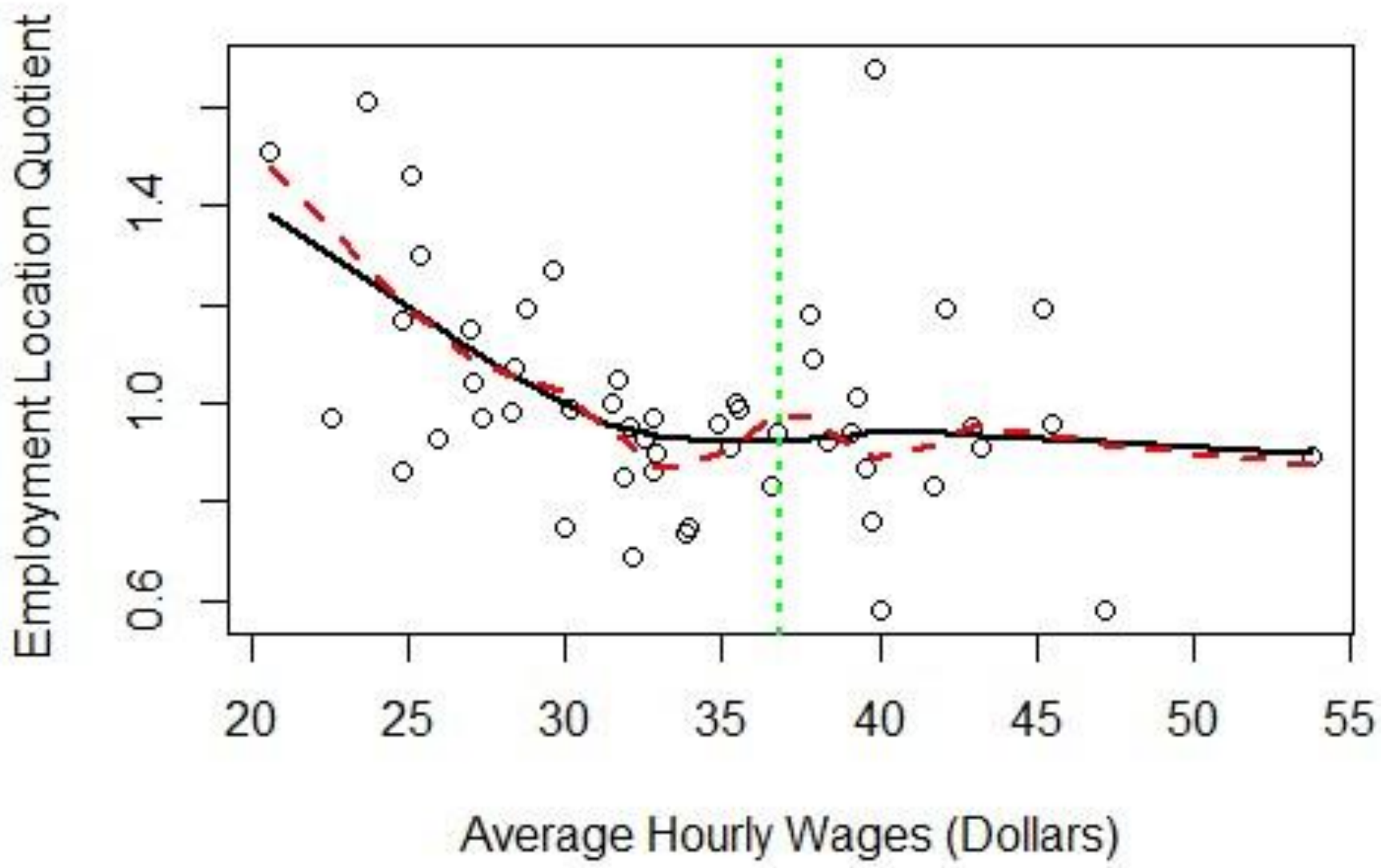

**Figure 3.2** This figure shows two scatterplot smooths based on the Police and Sheriff Patrol Officers data. The graph at the top is the scatterplot from the highest perceptual group in the linked micromaps plot in Figure 3.1. This includes a lowess curve created using a default smoother span of 2/3. The bottom graph explores the extent to which the curve may be sensitive to the choice of span. Like the one from `micromapST`, the bottom graph uses the lowess function (local linear fit) in base R. The black solid curve is the same one in the top graph, with a default span of 2/3. The dashed red line uses a span of 0.3. On the far-left side of the graph, the dashed red line is somewhat higher than the solid black line, reflecting the effects of some points that have relatively high LQ values. In addition, the dashed red line displays somewhat greater variability in the middle of the graph for units relatively close to the mean hourly wage (vertical green line). It is important to explore data sets by varying models and parameters to gain insights into the data and potential relationships to propose. RETURN

**Figure 3.3** This linked micromaps plot displays the location quotients (LQs) and wages for elementary school teachers of general education classes and special education. The states are sorted based on the LQ for special education teachers as shown in the first data column. Consequently, the states in each perceptual group will have similar location quotients. The next column contains scatterplots of the LQ for general education teachers versus the LQ for special education teachers. Adding a lowess smooth hints at a slight nonlinear relationship. Subject matter experts might explore deviations from this lowess curve. For instance, in the last perceptual group, MO and AL are above the line and might be potential outliers. It would be of interest to explore the extent to which the deviations might be attributable to regional differences within the state. The wages for the two categories of teachers are compared in the scatterplots displayed in the last column and are showing an approximate linear relationship. These scatterplots display three variables – wages (via the two axes) and employment of special education teachers (via the filled in circles). One interesting state is WV, which has the highest LQ for special education teachers, but whose wages are at the lowest end of the range. As with Figure 3.1, readers are encouraged to try different smoother spans with the lowess. See the files on the github site for example code. RETURN

Software Developers - Employment & Wages
Occupational Employment & Wage Statistics 2023
Two Ended Cumulative Maps
States Highlighted
U. S. States
Location Quotient Employment
Hourly Wage w/ 90% CI
Hourly Wage Percentiles
WA VA CA MA CO
UT NH MN NJ AZ
NC DC MD NY OR
RI CT GA DE KS
TX MI IL FL AL
Median for Sorted Panel
NE
OH WI ID PA MO
IA VT SC ME TN
NM SD MT OK AR
HI ND IN NV KY
MS WV WY LA AK
LQ
Dollars
Nat Mean $66
Dollar
Nat Median $64

**Figure 3.4** This figure shows how linked micromaps can be used to display many types of variability in an intuitive way. This graphic shows employment (sort variable) and wages for the Software Developers occupation. Data from all nine columns in Table 1 are visualized in this linked micromaps plot. Additional reference values relevant to each column are shown as vertical green lines to help assess distribution information such as location and variability as compared to national values. Note especially that the middle data column displays state-level mean wage estimates, accompanied by 90% confidence intervals, and thus conveys visually one important measure regarding variability. In addition, the rightmost column presents state-level percentiles for the wage data, thus conveying a second, complementary indication of variability in the available wage information. Including both data columns in the linked micromaps enables the exploration of both across-state patterns of mean wages and of heterogeneity of wages within states. RETURN